\documentclass[11pt]{article}

\usepackage{amsmath,amssymb}
\usepackage{graphicx}
\usepackage{epsfig}

\newcommand{\be}{\begin{equation}}
\newcommand{\ee}{\end{equation}}
\newcommand{\bea}{\begin{eqnarray}}
\newcommand{\eea}{\end{eqnarray}}
\newcommand{\ba}{\begin{array}}
\newcommand{\ea}{\end{array}}

\newcommand{\alp}{\alpha'}

\newcommand{\cH}{{\cal{H}}}

\newcommand{\fft}[2]{{\frac{#1}{#2}}}

\begin{document}

\begin{titlepage}
\begin{flushright}
\end{flushright}

\vfill

\begin{center}
{\large \bf
Drag force in SYM plasma with B field from AdS/CFT}
\vfill
{
Toshihiro Matsuo$^{*,}$\footnote{\tt tmatsuo@home.phy.ntnu.edu.tw},
}
{
Dan Tomino$^{*,}$\footnote{\tt dan@home.phy.ntnu.edu.tw}
}
and
{
Wen-Yu Wen$^{\dagger,}$\footnote{\tt steve.wen@gmail.com}
}
\vfill
$^{*}$ {\it
Department of Physics,
National Taiwan Normal University,\\
Taipei City, 116, Taiwan
}
\vskip 0.4truecm
$^{\dagger}$ {\it
Physics Department,
National Taiwan University,\\
Taipei, Taiwan, R. O. C.
}

\end{center}

\vfill
\begin{abstract}
We investigate drag force in a thermal plasma of N=4 super Yang-Mills theory via both fundamental and Dirichlet strings under the influence of non-zero NSNS $B$-field background.
In the description of AdS/CFT correspondence the endpoint of these strings correspondes to an external monopole or quark moving with a constant electromagnetic field.
We demonstrate how the configuration of string tail as well as the drag force obtains corrections in this background.

\end{abstract}
\vfill

\end{titlepage}

\setcounter{footnote}{0}


\section{Introduction}

At the experiments of Relativistic Heavy Ion Collision (RHIC),
collisions of gold nuclei at $200$ GeV per nucleon are about to
produce a strongly-coupled quark-gluon plasma (QGP), which behaves
like a nearly ideal fluid \cite{RHIC}.
While the perturbative calculation can not be fully trusted in this
strongly-coupled regime, there are increasing amount of interests
in calculation of hydrodynamical transport quantities via the use
of AdS/CFT correspondence \cite{hydro} \cite{jetq}.
It is hoped that this line of research will eventually make contact
with experiment result from RHIC.
As it was first proposed \cite{Maldacena:1997re}, the large $N$ limit
of ${\cal N}=4$ super Yang-Mills field theories in four dimensions
include in their Hilbert space a sector describing supergravity on
the $AdS_5\times S^5$ background.
The curvature of the sphere and the AdS space in units of $\sqrt{\alpha'}$ is proportional to $1/\sqrt{N}$, therefore the solutions can be trusted as long as $N$
is large.
In this paper, we are interesting in probing this strongly-coupled QGP by some {\sl theoretically} possible particles, in particular a heavy quark or a heavy monopole.
This corresponds respectively to dragging an attached fundamental or
Dirichlet string tail cross the whole AdS bulk.
Finite temperature configurations in the decoupled field theory
correspond to black hole configurations in AdS spacetimes.
In particular, there is a linear relation between the size of horizon
and Hawking temperature for large AdS black holes.
It is suggested by the holographic principle that properties of this
strong interaction for probe particles with a hot plasma of gluons
and quarks are in part or fully reflected on the dual gravitational
background.
Apart from the vacuum AdS, or correspondingly super conformal field theory on the boundary, introduction of black hole is hoped to better describe a realistic QCD model on the dual field theory side, though the correspondence is so far not fully satisfied yet.

A probe particle can be prepared by separating a pair of
particle-antiparticle in the deconfinement phase to far distance,
then it is enough to consider just one single particle with a string
tail all the way up to the black hole horizon. Description of the
other particle in the same pair is just a mirror to this. Energy put
to this single particle, if any, must either alter the kinetic
energy, or dissipated into the surrounding plasma via strong
interaction. In particular the latter can be holographically
described as dissipated energy being dumped into the black hole via
the string energy-momentum current.

Energy loss of a heavy quark moving through ${\cal N}=4$ super Yang-Mills thermal plasma has been extensively studied recently \cite{energy loss}.
In particular, the drag force is derived in the context of AdS/CFT to model the effective viscous interaction \cite{hkkky,gubser}, later it is generalized to a
rotational black hole or couple to dilaton field \cite{herzog,cacg1,cacg2}.  Recently drag force of a comoving quark-antiquark pair is also considered  in \cite{Chernicoff:2006hi}
.\\

In this note, we investigate the drag force via both fundamental
and Dirichlet strings under the influence of non-zero NSNS
$B$-field background. It may be interpreted as an constant
electromagnetic field on the D3-brane. We expect that it also
sheds some light to real QGP under an external electromagnetic
field. From the viewpoint of string theory, B-field is the gauge
field to which a string can couple. However it is not clear how
does it affects the movement of string endpoint on the AdS
boundary, because it couples to F1 and D1 strings in a different
way.  This paper is organized as follows. In section 2 we
calculate the drag force of an external monopole moving in a
thermal plasma with constant electromagnetic field in the ${\cal
N}=4$ super Yang-Mills theory. In section 3 we consider an
external quark in a {\sl noncommutative} super Yang-Mills theory
at finite temperature. In section 4, we give a summary and
discussion.

\section{Heavy monopole probe in QGP}

In this section, we consider a monopole which moves in constant
speed, and introduce a constant B-field of either electric-type $E$
or magnetic-type $H$ along the $x^1$ and $x^2$ direction. Because
only the field strength is involved in equations of motion, our
ansatz is still a good solution to supergravity. For the dual field
theory, this is the minimal setup to investigate on the B-field
correction. A simple relevant geometry is a AdS$_5 $ Schwarzschild
black hole with constant B-field,
\begin{eqnarray}\label{ads_BH}
&&ds^2={\cal H}^{-1/2}(-hdt^2+dx_1^2+dx_2^2+dx_3^2)+{\cal H}^{1/2}h^{-1}\frac{dr^2}{r^2} , \nonumber\\
&&B=Edt\wedge dx_1 + Hdx_1\wedge dx_2,\nonumber\\
&&h=1-\frac{r_H^4}{r^4}, \qquad {\cal H}=\frac{L^4}{r^4},
\label{adsbl}
\end{eqnarray}
where the AdS radius as well as $S^5$ is given by
$L^2=\sqrt{\lambda}\alpha'$, and the horizon is at $r=r_H$, which
is related to temperature by $T=r_H/{\pi L^2}$ for large black
hole.  We have introduced the NS-NS antisymmetric field $B_{01}=E$
and $B_{12}=H$. These $E$ and $B$ are constants.

In the AdS/CFT prescription, monopole corresponds to the endpoint
of an open D-string in the bulk supergravity. We may describe such a string attached to the monopole by
\begin{equation}\label{quark_motion}
x_1=v_1 t + \xi_1(r), \qquad x_2= v_2t+\xi_2(r)
\end{equation}
because we have introduced $B_{01}$ and $B_{12}$ only.    We let the string worldsheet $(\tau,\sigma)$ span along $t$ and
$r$ directions. For convenience we choose worldsheet
coordinates $\tau=t$ and $\sigma=r$.   As mentioned above, this can also be understood as departing a
pair of long-distance separated monopole-antimonopole.  While they
depart from each other, drag force is generated due to string
tension induced via the bulk geometry.

The D1 string action is described by the DBI action:
\begin{eqnarray}
&&S_{DBI}=-\frac{1}{2\pi\alpha'g_s}\int{d\tau
d\sigma\sqrt{-det(g+b)_{ab}}} .
\end{eqnarray}
The induced metric $g$ and $b$-field on
the D1 string worldsheet is given by
$det(g+b)_{ab}=det(G+B)_{\mu\nu}\partial_{a}X^{\mu}\partial_{b}X^{\nu}$,
where $a,b=\tau,\sigma$.  $G,B$ and $g,b$ respectively denote the
spacetime and induced worldsheet metric and B-field.  That is,
\begin{eqnarray}
&&g_{\tau\tau}= -{\cal H}^{-1/2}h + {\cal
H}^{-1/2}|\vec{v}|^2,\qquad  b_{\tau\tau}=0,\nonumber\\
&&g_{\tau\sigma}=g_{\sigma\tau}={\cal H}^{-1/2}\vec{v}\cdot
\vec{\xi}^{'},\qquad
b_{\tau\sigma}=-b_{\sigma\tau}=E{\xi_1^{'}}+H\vec{v}\times\vec{\xi}^{'},\nonumber\\
&&g_{\sigma\sigma}={\cal H}^{1/2}h^{-1}+{\cal
H}^{-1/2}|\vec{\xi}^{'}|^2, \qquad b_{\sigma\sigma}=0,
\end{eqnarray}
where two vectors on the $x_1$-$x_2$ plane, velocity
$\vec{v}=(v_1,v_2)$ and projected direction of string tail
$\vec{\xi^{'}}=(\xi_1^{'},\xi_2^{'})$, are in generic pointing to
different directions.  $\vec{v}$ and $\vec{\xi^{'}}$ are are
opposite directions if the monopole-antimonopole pair are departed
from each other, while perpendicular if they are parallel
transported.  In this paper we only consider the former case.

Therefore the Lagrangian density is given by square root of the
following determinant
\begin{eqnarray}
{\cal L}^2=-det(g+b)=1+\fft{h}{\cal
H}|\xi^{'}|^2-\frac{|\vec{v}|^2}{h}-\frac{|\vec{v}\times\vec{\xi}^{'}|^2}{\cal
H}-(E\xi_1^{'}+H\vec{v}\times\vec{\xi^{'}})^2 .
\end{eqnarray}

\subsection{Dragging Monopole in the electric field}

\begin{figure}
\center{
\includegraphics{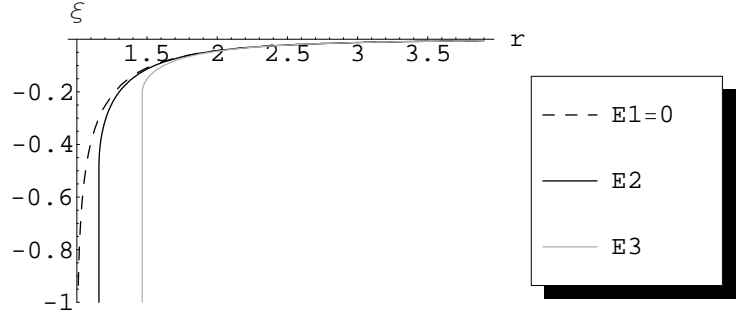}}
\caption{Configuration of string tail v.s. electric B-field
($E1<E2<E3$)}
\label{fig1}
\end{figure}

We now consider a monopole moving in a constant speed with constant electric field $E=B_{01}$.
To study an influence of $E$, we may choose the moving direction of monopole to be in the $x^1$ direction. In this case, the relevant trajectory
of the D1 string is more restricted than in (\ref{quark_motion})  and assumed to be
\begin{equation}\label{motion_electric}
x_1=vt+\xi_1(r),\qquad x_2=0 .
\end{equation}
Then
\begin{equation}
-det(g+b)=1-\frac{v^2}{h}+\alpha{\xi_1^{'}}^2,\qquad
\alpha=\frac{h}{\cal H}-E^2.
\end{equation}
With the constant conjugate $\pi_{\xi_1}$ calculated by
$\frac{\partial {\cal L}}{\partial \xi_1'}$, one obtains
\begin{equation}
\xi_1^{'}=\pm
\pi_{\xi_1}\sqrt{\frac{1-{v^2}/{h}}{\alpha^2-\alpha\pi_{\xi_1}^2}} .
\end{equation}
The positive sign is chosen for string tail moving behind the quark, i.e. $\xi'_1<0$.
Then the negative $\pi_{\xi_1}$ can be interpreted as spacetime momentum flow along $r$-direction on the worldsheet, from the boundary to the black hole.
We would like to remark that for the case $\pi_{\xi_1}>0$, which corresponds to momentum flow from black hole to the boundary, there will be $\xi_1'>0$ and string is moving ahead the quark.
This solution is in fact not physical because classically nothing can get out of black hole.
The reality condition for $\xi_1^{'}$ requires the denominator and numerator have a common root at the radius $r_v^4=\frac{r_H^4}{1-v^2}$, which gives rise to $\pi_{\xi_1}^2=\alpha(r_v)$.  After substituting back, we obtain
\begin{equation}
\xi_1^{'}=v\frac{L^2r_H^2}{r^4-r_H^4}\sqrt{\frac{1-a_eE^2}{1-b_eE^2}},\qquad
a_e=\frac{L^4}{r_H^4}\frac{1}{\gamma^2v^2},\qquad b_e=\frac{\cal
H}{h},
\end{equation}
where $\gamma=(1-v^2)^{-1/2}$ is the inverse Lorentz contraction
factor.
The reality condition for $\xi_1(r)$ implies a critical value $E_c=a_e^{-1/2}$ for $E$-field as well as a IR cut-off for $r$:
\begin{equation}
r>r_{IR}=(r_H^4+E^2L^4)^{1/4},\qquad  E<E_c .
\end{equation}
We summarize the result in Figure \ref{fig1} which shows the
configuration of string tail $\xi_1(r)$ into AdS bulk for
different strength of $E$.

To calculate the flow of energy dissipated into infrared along the
string, we first recall the conserved worldsheet current
associated with spacetime energy-momentum along $x_i$ is,
\begin{equation}
P^r_{x_i} \equiv \frac{1}{2\pi\alpha^{'}g_s}\frac{\delta{\cal
L}}{\delta\partial_r x_i} = -\frac{1}{2\pi\alpha^{'}g_s}\pi_{\xi_i} .
\end{equation}

For a particle with momentum $\vec{p}$ moving in a viscous medium
and subject to a driving force $\vec{f}$, we have
\begin{equation}\label{net force}
\dot{\vec{p}}=-\mu\vec{p}+\vec{f},
\end{equation}
where $\mu$ is the damping rate.  At the constant speed in our
case, we conclude the drag force is given by
\begin{equation}\label{drag_force_E}
-f_1=P^r_1=-\frac{1}{2\pi\alpha^{'}g_s}\sqrt{\gamma^2v^2\frac{r_H^4}{L^4}-E^2} .
\end{equation}

The square root shows its nonperturbative nature in the large $E$
limit, however it is also interesting to see the linearized result
for small $E$ expansion of (\ref{drag_force_E}),
\begin{equation}\label{energy_loss_E}
-f_1\simeq -\frac{1}{2\pi\alpha'g_s}\frac{r_H^2}{L^2}\gamma
v\left( 1-\frac{L^4}{2r_H^4}\gamma^{-2}v^{-2}E^2+{\cal
O}(E^4)\right).
\end{equation}
Notice that the correction terms are input as even power of $E$,
thanks to the determinant nature of DBI action.  As $E$ is
switched off, we reproduce the same result as shown in \cite{hkkky,gubser}. It is also instructive to rewrite the drag
force in terms of gauge theory parameters, i.e.
\begin{equation}\label{drag_force}
-f_1\simeq-\frac{\pi\sqrt{\lambda}}{2g_s}T^2\gamma
v\left(1-\frac{1}{2\pi^4\lambda{\alpha'}^2}T^{-4}\gamma^{-2}v^{-2}E^2
+{\cal O}(E^4) \right) .
\end{equation}

\begin{figure}
\center{
\includegraphics{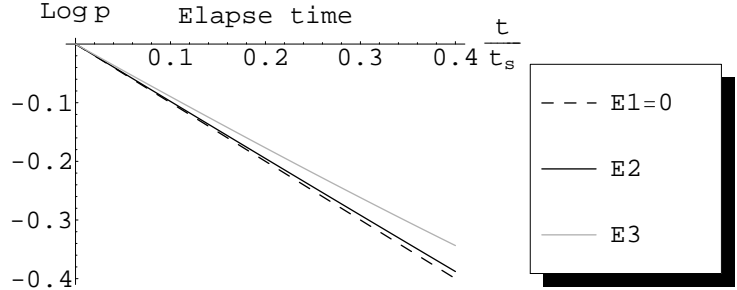}}
\caption{Elapse time v.s. B-field ($E1<E2<E3$)}
\label{fig2}
\end{figure}

On the other hand, we may consider a free monopole without driven
force.  Then equation (\ref{net force}) gives us the damping rate
or equivalently elapse time constant as inverse $\mu$.  In the
relativity limit, we have $p_1=\gamma m v$
\footnote{  Here we use relativistic dispersion relation $p=\gamma mv$ without thermal correction. This correction is expanded by
$\Delta m(T)/m$ with a function $\Delta m(T)$ \cite{hkkky}.  For heavy particle we can ignore it.}
for monopole mass $m$.
Here we plot (\ref{drag_force_E}) in Figure \ref{fig2} to show that the
elapse time constant against different strength of $E$.  %
For vanishing $E$, one reproduces the {\sl natural} relaxation time
constant
\begin{equation}
\tau_0^{-1}=\frac{2g_sm}{\pi\sqrt{\lambda}T^2},
\end{equation}
however, for nonzero $E$ elapse time constant is {\sl not} a
constant. We remark that the appearance of $g_sm$ implies that the
effective mass is the same as quark.  Therefore we expect this
gives us the elapse time constant of the same order as in the F1
string case.

\subsection{Dragging monopole in the magnetic field}
In this case, we consider the trajectory is given by
\begin{equation}
x_1=vt+\xi_1(r),\qquad x_2=\xi_2(r),
\end{equation}
with nonzero $H=B_{12}$.  Then
\begin{equation}
-det(g+b)=1-\frac{v^2}{h}+\frac{h}{\cal H}{\xi_1^{'}}^2 +\beta
{\xi_2^{'}}^2,\qquad \beta=\frac{h-v^2}{\cal H}-H^2v^2.
\end{equation}
The $\xi_i^{'}$ can be obtained by similar calculation as previous section,
\begin{eqnarray*}
\xi_1^{'2}&=&\pi_{\xi_1}^2\frac{ \left(1-\frac{v^2}{h}\right)\beta}
{ \frac{h}{{\cal H}}[(\pi_{\xi_1}^2-\frac{h}{{\cal
H}})(\pi_{\xi_2}^2-\beta)-\pi_{\xi_1}^2\pi_{\xi_2}^2]} ,
\\\\
\xi_2^{'2}&=&\pi_{\xi_2}^2\frac{\left(1-\frac{v^2}{h}\right)\frac{h}{{\cal H}}}
{\beta[(\pi_{\xi_1}^2-\frac{h}{{\cal H}})(\pi_{\xi_2}^2-\beta)-\pi_{\xi_1}^2\pi_{\xi_2}^2)]} ,
\end{eqnarray*}
for two conserved quantities $\pi_{\xi_1}$ and $\pi_{\xi_2}$. Then
the reality condition gives
\begin{equation}
\pi_{\xi_1}^2=\frac{r_H^4}{L^4}\frac{v^2}{1-v^2},\qquad
\pi_{\xi_2}^2=0,
\end{equation}
at the end, we have
\begin{eqnarray}
&&-f_1=-\frac{1}{2\pi\alpha'g_s}\frac{r_H^2}{L^2}\gamma v_1,
\nonumber\\
&&f_2=0 .
\end{eqnarray}

Notice that there is no drag force along $x_2$-direction and the
$B_{12}$ has no effect on the motion along $v_1$ at all.  This may
surprise us at first glance.  However, it could justify the
viscous property of drag force, which is simply against the
movement.  Therefore that $f_2=0$ in fact agrees with a vanishing
$v_2$.\\

Here we would like to remark about the case of ${\vec v}\perp E$ which we left in the previous subsection.
The analysis is similar to the one in this subsection.
We find that there are two solutions.
First solution gives $\pi_1=0$ therefore we have no drag force along the moving direction.
Second case gives $\pi_2=0$. In this case we have no effect of $E$.
Therefore the situation given in (\ref{motion_electric})  is sufficient to get relevant result.

\section{Heavy quark probe in QGP}
In this section we will calculate the drag force in hot plasma with large external B-field by using the gravity description for non-commutative supersymmetric Yang-Mills theory \cite{Maldacena:1999mh}
\footnote{Similar calculations on the same background have been done in \cite{Dhar:2000nj}, but at zero temperature.}%
.
We consider a fundamental string whose endpoints represent a quark and an anti-quark and use the quark as a probe in hot non-commutative SYM plasma.
The dynamics of the fundamental string is governed by Nambu-Goto action
\footnote{
The endpoints of string can couple to the B-field through the sigma model action.
Here we simply neglect the term because the coupling turns out to mealy introduce to the Lorentz force to the end point quark in the case of constant field.}
\begin{eqnarray}
S=-\frac{1}{2\pi \alp} \int d\tau d \sigma
\sqrt{-det(\eta^{\alpha \beta}
\partial_\alpha X^{\mu}\partial_\beta X^{\nu} G_{\mu\nu} )} .
\label{NGB}
\end{eqnarray}
Here the background metric $G_{\mu \nu}$ is deformed by a external B-field, which is given in the string frame \cite{Maldacena:1999mh};
\bea
ds^2&=&\cH^{-1/2}
\left[
-hdt^2+dx_1^2+k^{-1}(dx_2^2+dx_3^2)
\right]
+\cH^{1/2}h^{-1}dr^2+L^2 d\Omega_5^2 ,
\nonumber\\
&& B_{23}=B_{\infty}\frac{b^4r^4}{k}, \quad
B_{\infty}=\frac{\alp}{\theta},
\nonumber\\
&&e^{2\phi}=\hat{g}_s^2 k^{-1},
\nonumber\\
&&A_{01}=\frac{\theta}{\alpha'\hat{g}_s}\frac{r^4}{L^4} , \quad
F_{0123r}=\frac{1}{\hat{g}_sL^4}\frac{ \partial_{r}(r^4)}{k} ,
\nonumber\\
&& \cH=\frac{L^4}{r^4}, \quad h=1-\frac{r_H^4}{r^4} , \quad k=1+b^4r^4 ,
\quad b^2=\frac{\theta}{\alpha'L^2} ,
\label{mrback}
\eea
where $\hat{g}_s$ and $\theta$ are kept finite under the decoupling zero slope limit.
Here $\hat{g}_s$ is the string coupling constant, $\theta$ measures the non-commutativity on the D3-brane worldvolume and
$L^4/\alpha'^2=4\pi\hat{g}_sN$ corresponds to the 't Hooft coupling of the non-commutative SYM theory.
The background provides a supergravity dual description of non-commutative Yang-Mills theory at finite temperature
\footnote{
This non-commutative theory has an interpretation as the SYM with large and constant B field without non-commutativity\cite{Seiberg:1999vs}
.
However we do not consider such a commutative limit here.
}
.

\subsection{Dragging quark in non-commutative space}
The magnetic field here points to $x^1$-direction, and we have
rotational symmetry in the $x^2$-$x^3$ plane, thus we set a frame in
which a quark moving along the $x^2$-direction which is
perpendicular to the magnetic field. Equivalently we consider a
quark moving in the non-commutative  plane ($x^2$-$x^3$ plane)
\footnote{In the case of a quark moving parallel to the magnetic field or perpendicular to the non-commutative plane, the calculation is same as in the case of no B-field.}
.
The string worldsheet can be written
\bea
y(t,r)=v t + \zeta(r) .
\eea
Plugging this into the string action (\ref{NGB}), we obtain
\bea
S=-\frac{1}{2\pi \alp} \int dtdr \sqrt{
1-\frac{v^2}{hk}
+\frac{h}{\cH k}\zeta'^2 } .
\eea
One can easily find the string configuration in the $r$-direction has a derivative:
\bea
\zeta'(r)
=\pm \frac{\cH k}{h}\pi_\zeta \sqrt{\frac{hk-v^2}{hk-\cH k^2\pi_\zeta^2}}
\eea
where $\pi_\zeta$ is the canonical momenta conjugate to $\zeta$ and
is a constant.
The reality condition determines the constant as
\bea
\pi_\zeta=\frac{v}{L^2}\frac{r_v^2}{1+b^4r_v^4}
\eea
where
\bea
r_v^4=\frac{1}{2b^4}
\left[-(1-v^2-b^4r_H^4)+\sqrt{(1-v^2-b^4r_H^4)^2+4b^4r_H^4}\right].
\eea
The drag force is given by this constant as
\bea
-f_2 = P_2^r=-\frac{1}{2\pi \alp} \pi_\zeta.
\eea

It is interesting to see the behavior in the limit of small and large $br_H$,
which corresponds to the limit of small and large $\theta$ respectively.
In the case of small $br_H$, $r_v$ can be expanded as
\begin{eqnarray}
r_v^4=\frac{r_H^4}{1-v^2}\left(1-\frac{v^2}{(1-v^2)^2}b^4r_H^4\right)
+O(b^8r_H^8).
\end{eqnarray}
Thus we obtain
\begin{eqnarray}
-f_2= -\frac{1}{2\pi \alp}\frac{r_H^2}{L^2}\frac{v}{\sqrt{1-v^2}}
\left[1-\frac{2-v^2}{2(1-v^2)^2}b^4r_H^4+O(b^8r_H^8)\right] .
\label{rhexp}
\eea
In a non-relativistic approximation which is valid for small velocity, one can assume $p = mv$ and the drag force becomes
\bea
-f_2\simeq -\frac{\pi}{2}\sqrt{\lambda}T^2 \frac{p_2}{m}
\Bigg[1-\pi^4T^4 \lambda \theta^2+ O(\theta^4)\Bigg] ,
\end{eqnarray}
where we have used the relations $b^4r_H^4=\pi^4T^4\lambda\theta^2$.
The relaxation time derived from this becomes
\bea
\tau_0 = \frac{2m}{\pi \sqrt{\lambda} T^2}
\left(1+ \pi^4T^4 \lambda \theta^2+ O(\theta^4)
\right) ,
\eea
which shows the non-commutative nature makes the medium less viscous.

Even in relativistic velocity, equation (\ref{rhexp}) is valid if $b^4 r^4_H \ll (1-v^2)^2$.
We use relativistic dispersion relation
$p=\gamma mv,$ then the force behaves as
\begin{eqnarray}
-f_2\simeq-\frac{\pi}{2}\sqrt{\lambda}T^2\frac{p_2}{m}
\left[  1 - \frac{\pi^4T^4\lambda\theta^2}{2m^4}p_2^4  + O(\theta^4) \right].
\end{eqnarray}
Dissipation of energy becomes  milder than in lower momentum region.
The above analysis shows that as $\theta$ is turn off the result reduces to the one obtained in \cite{hkkky,gubser}.

On the other hand, for large $br_H$ or $\theta$ we have
\begin{eqnarray}
b^4r_v^4=b^4r_H^4\left(1+\frac{v^2}{b^4r_H^4}+ O(b^{-8}r_H^{-8})\right)
\end{eqnarray}
therefore
\begin{eqnarray}
-f_2&=&-\frac{1}{2\pi \alp}
\frac{v}{L^2}\frac{1}{b^4r_H^2}\left(1-\frac{2+v^2}{2b^4r_H^4}+O(b^{-8}r_H^{-8})\right)
\nonumber \\
&\simeq&-\frac{1}{2\pi^3T^2\sqrt{\lambda}\theta^2}\frac{p_2}{m}
\left[1-\frac{1}{\pi^4T^4\lambda\theta^2}
+O\left(\theta^{-4}\right)\right]
\label{rhexp2}
\end{eqnarray}
for small velocity.
For relativistic velocity and $b^4 r_H^4 \gg (1-v^2)^2$, the first line in large  $b^4r^4_H$ expansion (\ref{rhexp2}) is valid.
Using the relativistic dispersion  relation then we have
\begin{eqnarray}
-f_2 \simeq -\frac{1}{2\pi^3T^2\sqrt{\lambda}\theta^2}
 \left[ 1 - \frac{3}{2\pi^4T^4\lambda\theta^2}  + O(\theta^{-4})    \right].
\end{eqnarray}
In this limit we find constant friction.
It is interesting to see the dependence of the 't Hooft coupling and temperature is inverted relative to the small non-commutative case.

A comment concerning to the relation to the ordinary Yang-Mills theory
with magnetic field is in order.
The non-commutative theory on D-brane allows another interpretation
as ordinary theory with (large) constant electromagnetic field on the D-brane.
According to this interpretation, our results imply that the constant magnetic field affects thermal plasma to decrease the drag force  for both large and small velocity.
In other words, a large magnetic field may decrease the effective viscosity of QGP.\\

\section{Summary and discussion}
We have investigate the issue on the drag force in thermal SYM with B-field or NCSYM via AdS/CFT with appropriate backgrounds.
We have used a quark and a monopole as probes to measure the drag force
in the thermal medium.

For the monopole probe we have studied the problem with AdS$_5$
Schwartzchild black hole background with constant B-field and use
the DBI action as effective action of D1 brane whose endpoint is the
monopole to calculate the drag force. We considered separately the
two cases in which either a constant $B_{01}$ or $B_{12}$ is turned
on. In the case of turning on $B_{01}$, we evaluate the drag force
in the $x^1$ direction by taking the monopole moving along this
direction, whereas in the $B_{12}$ case the motion of the monopole
is perpendicular to the magnetic force, and we allow the drag force
span in both $x^1$ and $x^2$ directions. We found for electric case
the effects of the B-field reduce the drag force regardless of its
sign.  This suggests turning on electric B-field effectively weakens
the viscosity of the medium to a monopole. To our surprise, the
magnetic B-field do not have any effect on the drag force of a
monopole, which somehow reveals its nature of viscous force.

For the quark probe we use Nambu-Goto action describing a
fundamental string, whose endpoint is the quark, with the background
containing the effect of constant B-field which results in the
non-commutative super Yang-Mills theory or ordinary SYM theory with
constant field strength on the boundary.
We have studied $\theta$ and $\theta^{-1}$ expansions of
the drag force of quark probe with small and large velocity.
Again, the result implies
that the effect of the constant B-field or non-commutativity makes
the medium less viscous.

Thus we suggest that static magnetic field may decrease the
effective viscosity of QGP to a quark. The natural guess is that
even in the situation of real QCD a constant magnetic field
makes thermal plasma less viscous.

We leave several problems as future works.
One of them is introducing $B_{01}$ in Maldacena-Russo blakc hole background to investigate the effect of static electric field.
Such a background is already known as the supergravity dual of non-commutative open string (NCOS) theory.
The metric is given by \cite{Harmac:2000}
\begin{eqnarray}
\frac{ds^2}{\alpha'}&=&
f^{-1/2}G(-hdt^2+dx_1^2)+f^{-1/2}(dx_2^2+dx_3^2)
+\frac{1}{\alpha'^{2}}f^{1/2}(h^{-1}dr^2+r^2d\Omega^2),
\nonumber\\
&&f=1+\frac{\alpha'^{2}L^4}{r^4},\quad
G=\frac{r^4}{\alpha'^{2}L^4}f^{-1},\quad
h=1-\frac{r_H^4}{r^4}.
\end{eqnarray}
where the factor $G$ represents the effect of $B_{01}$.
We can easily calculate the drag force with ansazt:
$x^1=vt+\xi(r)$ and obtain
\begin{eqnarray}
-f_1=-\frac{1}{2\pi}
\sqrt{1+\frac{1}{1-v^2}(\pi T_H)^4\lambda}
\;\frac{v}{\sqrt{1-v^2}}(\pi T_H)^2\sqrt{\lambda}.
\end{eqnarray}
The effect of electric field introduce a square root factor
$\sqrt{1+\frac{1}{1-v^2}(\pi T_H)^4\lambda}$ thus increases drag force.
This is contrast to the case of magnetic field where  drag force decreases.
The quark moving in hot NCOS which is not a field theory.
It is interesting to identify the medium to be described by NCOS.\\

At last, It remains to study the monopole case by considering DBI action with Maldacena-Russo background.
To estimate jet-quenching parameters in the setup investigated in the present paper may also be an interesting problem as well.

\section*{Acknowledgements}
We are grateful to organizers of the summer activity hosted in the National Taiwan University for providing a chance to present a part of this work in the early stage, especially thank T.C.~Chan, H.Y.~Chen, P.M.~Ho, T.~Inami, H.C.~Kao, R.~Sasaki, and S.~Teraguchi for useful discussion.
T.M also would like to thank Feng-Li Lin for stimulus conversations.
We are supported in part by the Taiwan's National Science
Council under grants NSC95-2811-M-003-005(TM), NC95-2811-M-003-007(DT) and  NSC95-2811-M-002-013(WYW).




\end{document}